\documentclass[aps,pra,10pt,twocolumn,amsmath,amssymb,
longbibliography,superscriptaddress,nofootinbib]{revtex4-1}

\usepackage[colorlinks=true,citecolor=blue,linkcolor=blue]{hyperref}
\usepackage{graphicx}
\usepackage{amsmath,amsthm,amssymb}
\usepackage{bm,bbm,mathbbol}
\renewcommand\Im{\operatorname{Im}}
\renewcommand\Re{\operatorname{Re}}

\newcommand{\eq}[1]{Eq.~(\ref{eq:#1})}
\newcommand{\eqs}[2]{Eqs.~(\ref{eq:#1}) and~(\ref{eq:#2})}
\newcommand{\equ}[1]{Equation~(\ref{eq:#1})}

\newcommand{\fig}[1]{Fig.~\ref{fig:#1}}
\newcommand{\sect}[1]{Sec.~\ref{sec:#1}}

\def\beq{\begin{equation}}
\def\eeq{\end{equation}}
\def\bea{\begin{eqnarray}}
\def\eea{\end{eqnarray}}

\def\ket#1{\vert#1\rangle}

\def\ip#1#2{\langle#1\vert#2\rangle}
\def\me#1#2#3{\langle#1\vert#2\vert#3\rangle}

\def\E{{\mathcal E}}
\def\EE{{\bm{\mathcal E}}}
\def\ww{\omega}

\def\dk{[d\kk]}

\def\kk{{\bm k}}
\def\rr{{\bm r}}
\def\Aa{{\bm A}}
\def\RR{{\bm R}}
\def\OO{{\bm 0}}

\def\btau{{\bm\tau}}
\def\W{^{\rm (W)}}

\def\nn{\nonumber\\}

\def\AAA{\mathbbm{A}}
\def\rrr{\mathbbm{r}}
\def\vvv{\mathbbm{v}}
\def\www{\mathbbm{w}}
\def\aaa{\mathbbm{a}}
\def\bbb{\mathbbm{b}}
\def\DDD{\mathbb{\Delta}}

\def\la{\langle\kern-2.5pt\langle}
\def\ra{\rangle\kern-2.5pt\rangle}
\def\vt{\vert\kern-1.5pt\vert}

\def\doteq{\,\overset{\boldsymbol{.}}{=}\,}

\begin{document}

\title{{\it Ab initio} calculation of the shift photocurrent by
  Wannier interpolation}

\author{Julen Iba\~{n}ez-Azpiroz} \affiliation{Centro de F{\'i}sica de
  Materiales, Universidad del Pa{\'i}s Vasco (UPV/EHU), 20018 San
  Sebasti{\'a}n, Spain}

\author{Stepan S. Tsirkin} \affiliation{Centro de F{\'i}sica de
  Materiales, Universidad del Pa{\'i}s Vasco (UPV/EHU), 20018 San
  Sebasti{\'a}n, Spain}

\author{Ivo Souza} \affiliation{Centro de F{\'i}sica de Materiales,
  Universidad del Pa{\'i}s Vasco (UPV/EHU), 20018 San Sebasti{\'a}n,
  Spain} \affiliation{Ikerbasque Foundation, 48013 Bilbao, Spain}

\date{\today}

\begin{abstract}
  We describe and implement a first-principles algorithm based on
  maximally-localized Wannier functions for calculating the
  shift-current response of piezoelectric crystals in the
  independent-particle approximation. The proposed algorithm presents
  several advantages over existing ones, including full gauge
  invariance, low computational cost, and a correct treatment of the
  optical matrix elements with nonlocal pseudopotentials.
  Band-truncation errors are avoided by a careful formulation of
  $k\cdot p$ perturbation theory within the subspace of wannierized
  bands. The needed ingredients are the matrix elements of the
  Hamiltonian and of the position operator in the Wannier basis, which
  are readily available at the end of the wannierization step.  If the
  off-diagonal matrix elements of the position operator are discarded,
  our expressions reduce to the ones that have been used in recent
  tight-binding calculations of the shift current. We find that this
  ``diagonal'' approximation can introduce sizeable errors,
  highlighting the importance of carefully embedding the tight-binding
  model in real space for an accurate description of the charge
  transfer that gives rise to the shift current.
\end{abstract}
\maketitle

\section{Introduction}
\label{sec:introduction}

Under homogeneous illumination, noncentrosymmetric crystals exhibit
the bulk photovoltaic effect (BPVE), a nonlinear optical response that
consists in the generation of a photovoltage (open circuit) or
photocurrent (closed circuit) when light is absorbed via intrinsic or
extrinsic
processes~\cite{fridkin_bulk_2001,sturman-book92,ivchenko-book97}. Contrary
to the conventional photovoltaic effect in $p$-$n$ junctions, the BPVE
occurs in homogeneous systems, and the attained photovoltage is not
limited by the band gap of the material.  The BPVE comprises a
``circular'' part that changes sign with the helicity of light, and a
``linear'' part that also occurs with linearly-polarized or
unpolarized light.  The former is symmetry-allowed in the gyrotropic
crystal classes, and the latter in the piezoelectric
ones~\cite{fridkin_bulk_2001,sturman-book92,ivchenko-book97}.

The present work deals with the intrinsic contribution to the linear
BPVE due to interband absorption, known as ``shift current.''  This
phenomenom was intensively studied in the 60s and 70s, particularly in
ferroelectric oxides such as BaTiO$_3$~\cite{koch-fe76}. In recent
years it has attracted renewed interest in view of potential
applications in novel solar-cell
designs~\cite{butler-ees15,tan-cm16,cook-nc17}, and in connection with
topological insulators~\cite{tan-prl16,braun-nc16,bas-oe16} and Weyl
semimetals~\cite{osterhoudt-arxiv17,yang-arxiv17,zhang-arxiv18}.

In a simplified picture, the shift current arises from a coordinate
shift accompanying the photoexcitation of electrons from one band to
another. Like the intrinsic anomalous Hall
effect~\cite{nagaosa-rmp10}, the shift current originates from
interband velocity matrix elements, depending not only on their
magnitudes but also on their
phases~\cite{baltz-prb81,belinicher-jetp82,kristoffel-zpb82,sipe-prb00}.

Over the years, the understanding of the shift current has greatly
benefited from model
calculations~\cite{presting-pssb82,fregoso-prb17,cook-nc17,tan-cm16}.
Tight-binding models have been used to analyze various aspects of the
problem, including the possible correlation with electric
polarization, the role of virtual transitions, and the sensitivity to
the wave functions.  Recently, density-functional theory methods
started being employed to calculate the shift-current responsivity in
specific
materials~\cite{nastos-prb06,young-prl12,tan-cm16,rangel-prl17}. The
results are generally in good agreement with experimental
measurements, proving the predictive power of the {\it ab initio}
approach.

The first-principles evaluation of the shift current (and of other
nonlinear optical responses) is technically challenging, due to the
intricate form of the matrix elements
involved~\cite{baltz-prb81,belinicher-jetp82,kristoffel-zpb82,sipe-prb00}. Two
basic approaches have been devised.  One is to express those matrix
elements as an infinite sum over intermediate virtual
states~\cite{baltz-prb81,belinicher-jetp82,sipe-prb00}. In practice
this requires calculating a large number of unoccupied bands, to
minimize truncation errors~\cite{nastos-prb06,rangel-prl17}.
Alternatively, the matrix elements can be recast in terms of
derivatives with respect to the crystal momentum $\kk$ of the initial
and final band
states~\cite{baltz-prb81,belinicher-jetp82,kristoffel-zpb82,sipe-prb00}.
This strategy circumvents the summation over intermediate states, but
its practical implementation requires a careful treatment of the
derivatives on a finite $k$-point grid in order to retain gauge
invariance and handle degeneracies~\cite{young-prl12}.  Finally, it
has been found that the shift current tends to converge slowly with
respect to the number of \textit{k} points used for the Brillouin zone
(BZ) integration~\cite{rangel-prl17}.  All these factors render the
shift current more challenging and expensive to calculate than the
ordinary linear optical conductivity.

In this work, we develop an accurate and efficient {\it ab initio}
scheme for calculating the shift current and related nonlinear optical
responses in the independent-particle approximation. The proposed
methodology, based on localized Wannier
functions~\cite{marzari-rmp12}, is closely related to the Wannier
interpolation method of calculating to the Berry curvature and the
intrinsic anomalous Hall conductivity~\cite{wang-prb06}.  In essence,
it consists in evaluating the matrix elements by $k\cdot p$
perturbation theory within the subspace of wannierized bands. This
strategy inherits the practical advantages of the sum-over-states
approach in the complete space of Bloch eigenstates, but without
introducing truncations errors. In addition, it has a very low
computational cost thanks to the compact basis set.  We will comment
on the relation between our methodology and a recent proposal with
similar characteristics~\cite{wang-prb17}.

Our Wannier-interpolation scheme distinguishes itself in two
aspects. First, it provides a physically transparent connection to
tight-binding approaches~\cite{cook-nc17}. This is achieved by
adopting a phase convention for the Bloch sums that includes the
Wannier centers in the phase factors, such that the resulting
expressions cleanly separate into two parts: an ``internal'' part that
only depends on the Hamiltonian matrix elements and Wannier centers
(the only ingredients in a typical tight-binding calculation), and an
``external'' part containing the off-diagonal position matrix
elements.  We find that the latter can give a sizeable contribution to
the shift current; moreover, its inclusion removes an artificial
symmetry of the shift-current matrix elements in two-band
tight-binding models~\cite{cook-nc17}.  These findings highlight the
importance of carefully embedding the tight-binding model in real
space --~via the position matrix elements~-- when calculating the
shift current.  The other salient feature of our formulation is that
it is fully gauge invariant. This is in contrast to previous
Wannier-based schemes, where a parallel-transport gauge was assumed
when calculating the interband matrix
elements~\cite{wang-prb06,wang-prb17}.

The manuscript is organized as follows. In \sect{prelims} we provide
some background on the microscopic theory of the shift current.  In
\sect{wannier} we first review the Wannier-interpolation scheme for
calculating the energy bands and the interband dipole matrix elements;
the same interpolation approach is then applied to the generalized
derivative of the interband dipole matrix, completing the list of
ingredients needed for evaluating the shift current. The technical
details of our electronic-structure and Wannier-function calculations
are described in \sect{comp-details}, and the resulting shift-current
spectra of GaAs and monolayer GeS are presented and discussed in
\sect{results}. We provide some concluding remarks in \sect{summary},
and leave additional technical discussions to the appendices.

\section{Preliminaries}
\label{sec:prelims}

\subsection{Definitions and background}   
\label{sec:defs}

Our starting point is the formalism of Sipe and Shkrebtii for
calculating second-order interband optical responses of bulk crystals
within the independent-particle approximation~\cite{sipe-prb00}.  The
basic ingredients are the interband dipole matrix, and its
``generalized derivative'' with respect to the crystal momentum
$\kk$. They are given by
\beq
\label{eq:r}
r^a_{\kk nm}=(1-\delta_{nm})A^a_{\kk nm}
\eeq
and
\beq
\label{eq:gen-der}
r^{a;b}_{\kk nm}=\partial_b r^a_{\kk nm}
-i\left(A^b_{\kk nn}-A^b_{\kk mm}\right)r^a_{\kk nm}
\eeq
respectively, where
\beq
\label{eq:A}
A^a_{\kk nm}=i\ip{u_{\kk n}}{\partial_a u_{\kk m}}
\eeq
is the Berry connection matrix, where $\ket{u_{\kk m}}$ denotes the
cell-periodic part of a Bloch eigenstate and $\partial_a$ stands for
$\partial/\partial k_a$.

The three equations above define Hermitean matrices in the band
indices $n$ and $m$.  Importantly, the first two transform covariantly
under band-diagonal gauge transformations,
\beq
\label{eq:gauge-cov}
\ket{u_{n}}\rightarrow e^{i\beta_{n}}\ket{u_{n}}
\Rightarrow
\begin{cases}
r^a_{nm}\rightarrow
e^{i(\beta_{m}-\beta_{n})}r^a_{nm},\\
r^{a;b}_{nm}\rightarrow
e^{i(\beta_{m}-\beta_{n})}r^{a;b}_{nm},
\end{cases}
\eeq
where the subscript $\kk$ has been dropped for brevity.
As a result, the combination
\beq
\label{eq:I-abc}
I^{abc}_{mn}=r^b_{mn}r^{c;a}_{nm}
\eeq
appearing in \eq{sigma-abc} below is gauge invariant.

Consider a monochromatic electric field of the form
\beq
\EE(t)=\EE(\ww)e^{-i\ww t}+\EE(-\ww)e^{i\ww t},
\eeq
with $\EE(-\ww)=\EE^*(\ww)$. Phenomenologically, the {\it dc}
photocurrent density from the linear BPVE
reads~\cite{fridkin_bulk_2001,sturman-book92,ivchenko-book97}
\beq
\label{eq:lbpve}
j^a=2\sigma^{abc}(0;\ww,-\ww)\Re\left[\E_b(\ww)\E_c(-\ww)\right].
\eeq
The third-rank response tensor is symmetric under
$b\leftrightarrow c$, and transforms like the piezoelectric tensor.
According to Eqs.~(38) and~(41) in Ref.~\onlinecite{sipe-prb00}, the
interband (shift-current) part of the response is given by
\bea
\label{eq:sigma-abc}
\sigma^{abc}(0;\ww,-\ww)&=&-\frac{i\pi e^3}{4\hbar^2}
\int\dk\sum_{n,m}f_{nm}
\left(I^{abc}_{mn}+I^{acb}_{mn}\right)\nn
&\times&\left[\delta(\omega_{mn}-\omega)+\delta(\omega_{nm}-\omega)\right].
\eea
Here $f_{nm}=f_n-f_m$ and $\hbar\omega_{nm}=E_m-E_n$ are differences
between occupation factors and band energies, respectively, and the
integral is over the first BZ, with $\dk=d^dk/(2\pi)^d$ in $d$
dimensions. Because $I^{abc}_{mn}$ is Hermitean, the right-hand-side
of \eq{sigma-abc} is real. Its transformation properties under
inversion and time-reversal symmetry are summarized in
Appendix~\ref{app:symm}.

For comparison, we also calculate the joint density of states (JDOS)
per crystal cell,
\beq
\label{eq:jdos}
D_{\rm joint}(\ww)=\frac{v_c}{\hbar}\int\dk\,\sum_{n,m}\,
  f_{nm}\delta(\omega_{mn}-\omega) 
\eeq
($v_c$ is the cell volume), and the interband contribution to the
absorptive (abs) part of the dielectric function~\cite{sipe-prb00},
\beq
\label{eq:dielectric}
\epsilon^{ab}_{\rm abs}(\ww)=\frac{i\pi e^2}{\hbar}
\int\dk\sum_{n,m}f_{nm}r^a_{nm}r^b_{mn}
\delta(\omega_{mn}-\omega).
\eeq
In nonmagnetic crystals $\epsilon^{ab}_{\rm abs}$ is purely imaginary
and symmetric, and we report values for
$\Im\epsilon^{ab}_{\rm r}=\Im\epsilon^{ab}_{\rm abs}/\epsilon_0$, the
imaginary part of the relative permittivity.

\subsection{Sum rule for the generalized derivative}
\label{sec:sum-rule}

The matrix elements $r^a_{nm}$ and $r^{a;b}_{nm}$ appearing in
\eq{sigma-abc} satisfy the identities
\beq
\label{eq:r-v}
r^a_{nm}=\frac{v^a_{nm}}{i\ww_{nm}}\quad(m\not= n)
\eeq
and
\bea
\label{eq:sipe-sum-rule}
r^{a;b}_{nm}&=&\frac{i}{\ww_{nm}}
\Bigg[
\frac{v^a_{nm}\Delta^b_{nm}+v^b_{nm}\Delta^a_{nm}}{\ww_{nm}}
-w^{ab}_{nm}\nn
&+&\sum_{p\not= n,m}\,
\left(
  \frac{v^a_{np}v^b_{pm}}{\ww_{pm}}
  -\frac{v^b_{np}v^a_{pm}}{\ww_{np}}
\right)
\Bigg]\,
(m\not= n),
\eea
where
\begin{subequations}
\label{eq:defs-1}
\begin{align}
\label{eq:v}
v^a_{nm}&=\frac{1}{\hbar}\me{u_n}{\partial_a \hat H}{u_m},\\
\label{eq:Delta}
\Delta^a_{nm}&=\partial_a\ww_{nm}=v^a_{nn}-v^a_{mm},\\
\label{eq:w}
w^{ab}_{nm}&=\frac{1}{\hbar}\me{u_n}{\partial^2_{ab} \hat H}{u_m}.
\end{align}
\end{subequations}
\equ{r-v} can be obtained by differentiating the identity
$\me{u_n}{\hat H}{u_m}=E_n\delta_{nm}$ with respect to $k_a$ for
$m\not =n$. Differentiating once more with respect to $k_b$ and
inserting a complete set of states yields the sum rule in
\eq{sipe-sum-rule}~\cite{sipe-prb00,cook-nc17}.  For Hamiltonians of
the form $\hat H_\kk=(\hat{\bm p}+\hbar\kk)^2/2m_e+V(\rr)$, the term
$w^{ab}_{nm}$ therein has no off-diagonal components and does not
contribute to the sum rule.  That term should however be included in
tight-binding calculations~\cite{cook-nc17}, and in first-principles
calculations with nonlocal pseudopotentials~\cite{wang-prb17}.

\equ{sipe-sum-rule} has been used in {\it ab initio} calculations of
the shift current~\cite{nastos-prb06,rangel-prl17}, with a truncated
summation over intermediate states~$p\not= n,m$.  An exact
(truncation-free) expression for $r^{a;b}_{nm}$ that only requires
summing over a {\it finite} number of wannierized bands,
\eq{gen-der-wannier} below, constitutes a central result of the
present work.

\section{Wannier interpolation scheme}
\label{sec:wannier}

The needed quantities for calculating the shift-current response from
\eq{sigma-abc} are the energy eigenvalues, and the matrix elements
$r^a_{nm}$ and $r^{a;b}_{nm}$ defined by \eqs{r}{gen-der}.  In this
section we describe how to evaluate each of them in a Wannier-function
basis.

Consider a set of $M$ well-localized Wannier functions per cell
$w_j(\rr-\RR)=\ip{\rr}{\RR j}$ spanning the initial and final states
involved in interband absorption processes up to some desired
frequency~$\ww$.  (In practice we shall construct them by
post-processing a first-principles calculation, using the method of
maximally-localized Wannier
functions~\cite{marzari-prb97,souza-prb01}.)  Starting from these
orbitals, we define a set of Blochlike basis states as
\beq
\label{eq:u-w}
\ket{u\W_{\kk j}}=
\sum_\RR\,e^{-i\kk\cdot(\hat\rr-\RR-\btau_j)}\ket{\RR j},
\eeq
where the superscript (W) stands for ``Wannier
gauge''~\cite{wang-prb06}.  Note that at variance with
Ref.~\onlinecite{wang-prb06}, we have chosen to include the Wannier
center
\beq
\label{eq:wann-center}
\btau_j=\me{\OO j}{\hat \rr}{\OO j}
\eeq
in the phase factor of \eq{u-w}.  This phase convention, often used in
tight-binding calculations, is the most natural one for expressing the
Berry connection and related geometric quantities in reciprocal
space~\cite{pythtb}.

\subsection{Energy eigenvalues}
\label{sec:eig}

The matrix elements of the first-principles Hamiltonian
$\hat H_\kk=e^{-i\kk\cdot\hat\rr}\hat He^{i\kk\cdot\hat\rr}$ between
the Blochlike states~(\ref{eq:u-w}) read
\bea
\label{eq:H-w}
H\W_{\kk ij}&=&
\me{u\W_{\kk i}}{\hat H_\kk}{u\W_{\kk j}}\nn
&=&\sum_\RR\,e^{i\kk\cdot(\RR+\btau_j-\btau_i)}
\me{\OO i}{\hat H}{\RR j}.
\eea
Diagonalization of this $M\times M$ matrix yields the
Wannier-interpolated energy eigenvalues,
\beq
\label{eq:eig}
\left(U^\dagger_\kk H\W_\kk U_\kk\right)_{nm}=E_{\kk n}\delta_{nm},
\eeq
where $U_\kk$ is the unitary matrix taking from the Wannier gauge to
the Hamiltonian gauge.  This Slater-Koster type of interpolation, with
the Wannier functions acting as an orthogonal tight-binding basis, has
been shown in practice to provide a smooth $k$-space interpolation of
the {\it ab initio} eigenvalues. (With disentangled Wannier functions,
the interpolation is faithful only within the so-called ``inner'' or
``frozen'' energy window~\cite{souza-prb01}.)

\subsection{Berry connection and interband dipole}
\label{sec:r}

The same interpolation strategy can be applied to other $k$-dependent
quantities. In particular, the Hamiltonian-gauge Bloch states
\beq
\label{eq:u-H}
\ket{u_{\kk n}}=\sum_{j=1}^M\,\ket{u\W_{\kk j}}U_{\kk jn}
\eeq
interpolate the {\it ab initio} Bloch eigenstates, allowing to treat
wavefunction-derived quantities.

As a first example, consider the Berry connection matrix defined by
\eq{A}. Inserting the above expression for $\ket{u_{\kk n}}$ in that
equation yields~\cite{wang-prb06}
\begin{subequations}
\label{eq:A-wannier}
\begin{align}
\label{eq:A-decomp}
A^a_{nm}&=\AAA^a_{nm}+\aaa^a_{nm},\\
\label{eq:AAA}
\AAA^a_{nm}&=i\left(U^\dagger\partial_a U\right)_{nm},\\
\label{eq:aaa}
\aaa^a_{nm}&=\left(U^\dagger A\W_a U\right)_{nm},
\end{align}
\end{subequations}
where $A\W_a$ in \eq{aaa} denotes a Cartesian component of the Berry
connection matrix in the Wannier gauge,
\bea
\label{eq:A-w}
\Aa\W_{\kk ij}&=&
i\ip{u\W_{\kk i}}{\partial_\kk u\W_{\kk j}}\nn
&=&\sum_\RR\,e^{i\kk\cdot(\RR+\btau_j-\btau_i)}
\me{\OO i}{\hat \rr-\btau_j}{\RR j}.
\eea

The term $\AAA^a_{nm}$ in \eq{A-wannier} carries the interpretation of
a Berry connection for the eigenvectors of $H\W$ (the column vectors
of $U$). Introducing the notation $\vt u_n\ra$ for those
vectors,\footnote{When the Wannier centers are included in the phase
  factors of the Bloch sums as in \eq{u-w}, the eigenvectors of $H\W$
  can be thought of as tight-binding analogues of the {\it
    cell-periodic} Bloch states, hence the notation $\vt u_n\ra$. The
  fact that Berry-phase-type quantities are defined in terms of the
  cell-periodic Bloch states is the reason why that phase convention
  is the most natural one for dealing with such quantities in
  tight-binding~\cite{pythtb}.}  \eq{AAA} becomes
$\AAA^a_{nm}=i\la u_n\vt \partial_a u_m\ra$. This is the ``internal''
Berry connection for the tight-binding model defined by \eq{H-w} in
terms of the Hamiltonian matrix elements and Wannier centers.

The extra term $\aaa^a_{nm}$ in \eq{A-wannier} arises from
off-diagonal matrix elements of the position operator in the Wannier
basis, as can be seen by inspecting the matrix element in \eq{A-w}
together with \eq{wann-center}.  In tight-binding formulations, it is
customary to postulate a diagonal representation for
$\hat\rr$~\cite{graf-prb95,bennetto-prb96,paul-prb03,pythtb,boykin-ejp10},
\beq
\label{eq:diag}
\me{\OO i}{\hat \rr}{\RR j}\doteq
\btau_i\delta_{\RR,\OO}\delta_{ji},
\eeq
where we have introduced the symbol ``$\overset{\boldsymbol{.}}{=}$''
to denote equalities that only hold only within this ``diagonal
tight-binding approximation'' (diagonal TBA). Thus, $\aaa^a_{nm}$ is
the part of the Berry connection matrix $A^a_{nm}$ that is discarded
when making the diagonal TBA, and we will refer to it as the
``external'' part.

For the interband dipole matrix of \eq{r} we get
\beq
\label{eq:r-wannier}
r^a_{nm}=
\begin{cases}
\rrr^a_{nm}+\aaa^a_{nm}&\text{ if }m\not=n\\
0&\text{ if }m=n
\end{cases},
\eeq
where
\begin{subequations}
\label{eq:rrr-vvv}
\begin{align}
\label{eq:rrr}
\rrr^a_{nm}&= (1-\delta_{nm})\AAA^a_{nm}=
\begin{cases}
\displaystyle
\frac{\vvv^a_{nm}}{i\ww_{nm}}&
\text{ if }m\not= n\\
0&\text{ if }m=n
\end{cases},\\
\label{eq:vvv}
\vvv^a_{nm}&=\frac{1}{\hbar}
\left[U^\dagger\left(\partial_a H\W\right) U\right]_{nm},
\end{align}
\end{subequations}
with $\partial_a H\W$ obtained by differenting the right-hand-side of
\eq{H-w}.  \equ{rrr} is the ``internal'' counterpart of \eq{r-v} for
$r^a_{nm}$. It can be derived in a similar manner, by differentiating
\eq{eig} with $m\not= n$.

\subsection{Generalized derivative of the interband dipole}

The energy eigenvalues and interband dipole matrix elements $r^a_{nm}$
are the only ingredients entering \eq{dielectric} for the dielectric
function, which has been previously evaluated by Wannier
interpolation~\cite{yates-prb07}.  \equ{sigma-abc} for the shift
current contains in addition the generalized derivative
$r^{a;b}_{nm}$, and in the following we describe how to evaluate it
within the same framework.

\subsubsection{Useful definitions and identities}
\label{sec:idents}

Our strategy will be to evaluate \eq{gen-der} for $r^{a;b}_{nm}$
starting from \eqs{A-wannier}{r-wannier} for $A^a_{nm}$ and
$r^a_{nm}$, respectively. Inspection of those equations reveals that
we need to differentiate with respect to $k_b$ the matrices
$\vvv^a_{nm}$ and $\aaa^a_{nm}$. Noting that both of them are of the
form
\beq
\overline{\cal O}= U^\dagger{\cal O}\W U
\eeq
and using the identity
\beq
\label{eq:delU}
\partial_b U=-iU\AAA^b,
\eeq
we find
\beq
\label{eq:dbObar}
\partial_b\overline{\cal O}=
U^\dagger\left(\partial_b{\cal O}\W\right)U
+i\left[\AAA^b,\overline{\cal O}\right].
\eeq
Writing $\AAA^b_{nm}$ in the commutator as
$\delta_{nm}\AAA^b_{nn}+\rrr^b_{nm}$ and then expanding
$\left[\rrr^b,\overline{\cal O}\right]$ as a sum over states yields
\begin{widetext}
\beq
\label{eq:der}
\partial_b\left(\overline{\cal O}\right)_{nm}=
\left[U^\dagger\left(\partial_b{\cal O}\W\right)U\right]_{nm}
-i\left(\overline{\cal O}_{nn}
-\overline{\cal O}_{mm}\right)\rrr^b_{nm}
+i\sum_{p\not= n,m}^M
\left(
  \rrr^b_{np}\overline{\cal O}_{pm}-
  \overline{\cal O}_{np}\rrr^b_{pm}
\right)
+i\left(\AAA^b_{nn}-\AAA^b_{mm}\right)\overline{\cal O}_{nm},
\eeq
\end{widetext}
where the contribution from intermediate states $p\not=n,m$ has been
separated out.  

We find it convenient to define an ``internal generalized derivative''
of the matrix $\overline{\cal O}$ in analogy with \eq{gen-der},
\beq
\label{eq:gen-der-internal-def}
\left(\overline{\cal O}\right)^{;b}_{nm}=
\partial_b\left(\overline{\cal O}\right)_{nm}
-i\left(\AAA^b_{nn}-\AAA^b_{mm}\right)\overline{\cal O}_{nm}.
\eeq
Note that this is equal to the sum of the first three terms in
\eq{der}.  Before proceeding, let us also define the following
internal quantities in analogy with \eq{defs-1},
\begin{subequations}
\label{eq:defs-2}
\begin{align}
\label{eq:bbb}
\bbb^{ab}_{nm}&=
\left[U^\dagger \left(\partial_b A\W_a\right) U\right]_{nm},\\
\label{eq:DDD}
\DDD^a_{nm}&=\vvv^a_{nn}-\vvv^a_{mm},\\
\label{eq:www}
\www^{ab}_{nm}&=\frac{1}{\hbar}
\left[
       U^\dagger\left(\partial^2_{ab} H\W\right) U
\right]_{nm}.
\end{align}
\end{subequations}

\subsubsection{Derivation}
\label{sec:der}

We begin by differentiating the term $\rrr^a_{nm}$ in \eq{r-wannier}
for $r^a_{nm}$. From \eq{rrr} we get
\beq
\partial_b\rrr^a_{nm}=\frac{i}{\ww_{nm}^2}\vvv^a_{nm}\DDD^b_{nm}
-\frac{i}{\ww_{nm}}\partial_b\vvv^a_{nm}\quad(m\not= n).
\eeq
Evaluating $\partial_b\vvv^a_{nm}$ with the help of \eq{der} and
expressing the result in the form of \eq{gen-der-internal-def},
\beq
\label{eq:del-r}
\partial_b\rrr^a_{nm}=\rrr^{a;b}_{nm}
+i\left(\AAA^b_{nn}-\AAA^b_{mm}\right)\rrr^a_{nm}\quad(m\not= n),
\eeq
we find
\bea
\label{eq:gen-der-internal}
\rrr^{a;b}_{nm}&=&\frac{i}{\ww_{nm}}
\Bigg[
\frac{\vvv^a_{nm}\DDD^b_{nm}+\vvv^b_{nm}\DDD^a_{nm}}{\ww_{nm}}
-\www^{ab}_{nm}\nn
&+&\sum_{p\not= n,m}^M\,
\left(
  \frac{\vvv^a_{np}\vvv^b_{pm}}{\ww_{pm}}
  -\frac{\vvv^b_{np}\vvv^a_{pm}}{\ww_{np}}
\right)
\Bigg]\,(m\not= n).
\eea
This is the internal counterpart of the sum
rule~(\ref{eq:sipe-sum-rule}), written in terms of the tight-binding
eigenvectors, eigenvalues, and Hamiltonian, instead of the {\it ab
  initio} ones.

The same procedure can be used to differentiate the term $\aaa^a_{nm}$
in \eq{r-wannier}, given by \eq{aaa}.  The result is
\beq
\label{eq:del-a}
\partial_b\aaa^a_{nm}=\aaa^{a;b}_{nm}
+i\left(\AAA^b_{nn}-\AAA^b_{mm}\right)\aaa^a_{nm},
\eeq
where
\bea
\label{eq:gen-der-internal-a}
\aaa^{a;b}_{nm}&=&\bbb^{ab}_{nm}-
\left(\aaa^a_{nn}-\aaa^a_{mm}\right)\frac{\vvv^b_{nm}}{\ww_{nm}}\nn
&+&
\sum_{p\not= n,m}^M
\left(
  \frac{\vvv^b_{np}\aaa^a_{pm}}{\ww_{np}}-
  \frac{\aaa^a_{np}\vvv^b_{pm}}{\ww_{pm}}
\right)\,(m\not=n).
\eea

Adding $\partial_b\rrr^a_{nm}$ and $\partial_b\aaa^a_{nm}$ from
\eqs{del-r}{del-a} to form $\partial_b r^a_{nm}$, and then subtracting
the amount $i\left(A^b_{nn}-A^b_{mm}\right)r^a_{nm}$ in the form
\beq
i\left(
  \AAA^b_{nn}+\aaa^b_{nn}
  -\AAA^b_{mm}-\aaa^b_{mm}
\right)\left(\rrr^a_{nm}+\aaa^a_{nm}\right)
\eeq
to obtain $r^{a;b}_{nm}$ as per \eq{gen-der}, we arrive at
\bea
\label{eq:gen-der-wannier}
r^{a;b}_{nm}&=&\rrr^{a;b}_{nm}+\aaa^{a;b}_{nm}
-\left(\aaa^b_{nn}-\aaa^b_{mm}\right)\frac{\vvv^a_{nm}}{\ww_{nm}}\nn
&-&i\left(\aaa^b_{nn}-\aaa^b_{mm}\right)\aaa^a_{nm}
\quad (m\not=n).
\eea
This expression for the generalized derivative in the Wannier
representation is a central result of the present work. An alternative
expression that is equally valid was obtained in
Ref.~\onlinecite{wang-prb17}, and the precise relation between the two
formulations is established in Appendix~\ref{app:comparison}.

\subsection{Discussion}

\subsubsection{Summary of the interpolation algorithm}

To summarize, the response tensor $\sigma^{abc}(0;\ww,-\ww)$ is given
by \eq{sigma-abc} in terms of the energy eigenvalues and of the matrix
elements $I^{abc}_{mn}$ defined by \eq{I-abc}.  At each~$\kk$, the
former are interpolated using \eq{eig}, and the latter using
\eqs{r-wannier}{gen-der-wannier} for $r^a_{nm}$ and $r^{a;b}_{nm}$,
respectively. These equations depend on a small number of ingredients:
the matrices $H\W$ [\eq{H-w}] and $A\W$ [\eq{A-w}], their first and
second mixed derivatives with respect to $k_a$ and $k_b$, and the
unitary matrix $U$ that diagonalizes $H\W$.  The needed real-space
matrix elements, $\me{\OO n}{\hat H}{\RR m}$ and
$\me{\OO n}{\hat \rr}{\RR m}$, can be evaluated as described in
Ref.~\onlinecite{wang-prb06}.

\subsubsection{Independence of the  Berry connection matrix
on the choice of phase convention for the Bloch sums}

It is well known that the tight-binding expression for an operator
depends on the phase convention used for the Bloch
sums~\cite{bena-njp09}. Let us discuss how this plays out for the
Berry connection matrix (similar remarks apply to the interband dipole
matrix and its generalized derivative).

The phase convention we have adopted in this work is that of \eq{u-w}.
The other commonly used convention is to drop $\btau_j$ from that
equation~\cite{pythtb,bena-njp09}, in which case the Berry connection
matrix is still given by \eq{A-wannier} but $\btau_i$ and $\btau_j$
should be removed from \eqs{H-w}{A-w}. As a result, the term
$\AAA^a_{nm}$ in \eq{A-wannier} becomes a function of the Hamiltonian
matrix elements only and not of the Wannier centers, whose
contributions to the Berry connection are absorbed by
$\aaa^a_{nm}$. The total Berry connection $A^a_{nm}$ remains the same
as before, but the term $\aaa^a_{nm}$ is now nonzero under the
diagonal TBA of \eq{diag}.

\subsubsection{Gauge covariance of the generalized derivative}

Although \eq{gen-der} for $r^{a;b}_{nm}$ is gauge covariant in the
sense of \eq{gauge-cov}, its individual terms are not, leading to
numerical difficulties. Instead, the individual terms in the
Wannier-based expression~(\ref{eq:gen-der-wannier}) for $r^{a;b}_{nm}$
transform covariantly under band-diagonal gauge transformations.  As a
result, its numerical implementation is very robust.

Contrary to Ref.~\onlinecite{wang-prb17}, we did not impose the
parallel-transport condition $\AAA^b_{nn}=0$ in our derivation of a
Wannier-based expression for $r^{a;b}_{nm}$. The gauge-dependent
quantities $\AAA^b_{nn}$ appear in intermediate steps of our
derivation, only to drop out in the final step leading to
\eq{gen-der-wannier}.  (A parallel-transport gauge was also assumed in
Ref.~\onlinecite{wang-prb06} when deriving a Wannier-based expression
for the Berry curvature, and in Appendix~\ref{app:curv} we indicate
how to remove that unnecessary assumption.)

\subsubsection{ Generalized derivative versus the effective-mass sum
  rule: The role of position matrix elements }

As remarked in \sect{sum-rule}, \eq{sipe-sum-rule} for $r^{a;b}_{nm}$
follows from differentiating the identity
$\me{u_n}{\hat H}{u_m}=E_n\delta_{nm}$ once with respect to $k_a$ and
once with respect to $k_b$, for $m\not= n$. Doing so for $m=n$ yields
the effective-mass sum rule.

For tight-binding models with a finite number of bands, the
effective-mass sum rule can be formulated exactly. The modified
sum-rule expression, which only depends on the Hamiltonian matrix
elements, includes an intraband term $\www^{ab}_{nn}$ given by
\eq{www}~\cite{graf-prb95,boykin-prb95,yates-prb07}.
 
The effect of the basis truncation on the calculation of nonlinear
optical responses has been the subject of several recent
investigations~\cite{cook-nc17,ventura-prb17,taghizadeh-prb17}. In
particular, it was suggested in Ref.~\onlinecite{cook-nc17} that
\eq{gen-der-internal} for $\rrr^{a;b}_{nm}$, which includes an
interband term $\www^{ab}_{nm}$, is the correct expression for
$r^{a;b}_{nm}$ in tight-binding models.  In fact, that expression only
accounts for part of the wavefunction dependence of $r^{a;b}_{nm}$,
via the diagonal position matrix elements.  The full expression,
\eq{gen-der-wannier}, has additional terms that depend on the
off-diagonal position matrix elements. Those should be included in
order to completely describe the wavefunction dependence, and to
render the result independent of the choice of Wannier basis
orbitals~\cite{paul-prb03}.

In the diagonal TBA of \eq{diag}, \eqs{r-wannier}{gen-der-wannier} for
$r^a_{nm}$ and $r^{a;b}_{nm}$ reduce to their internal terms,
$r^a_{nm}\doteq \rrr^a_{nm}$ and $r^{a;b}_{nm}\doteq \rrr^{a;b}_{nm}$.
In this approximation the shift current only depends on the
Hamiltonian matrix elements and on the Wannier centers, and a strong
dependence on the latter was found in Ref.~\onlinecite{cook-nc17}. As
we will see in \sect{results} (and also noted in
Ref.~\onlinecite{wang-prb17}), the additional contributions from
off-diagonal position matrix elements can modify appreciably the
calculated shift-current spectrum.

\subsubsection{The two-band limit}

The shift-current response of two-band tight-binding models has been
considered in Refs.~\onlinecite{presting-pssb82,cook-nc17}.  In that
limit the three-band terms in \eq{gen-der-wannier} (those containing
intermediate states) vanish identically, and $r^{a;b}_{nm}$ is
completely specified by the two-band terms, which pick up the missing
contributions (the importance of the $\www^{ab}_{nm}$ term in this
regard was emphasized in Ref.~\onlinecite{cook-nc17}). It appears to
have gone unnoticed that the diagonal TBA introduces a qualitative
error for two-band models, as we now discuss.

In the diagonal TBA, \eq{gen-der-wannier} for a two-band model reduces
to the first two terms in \eq{gen-der-internal},
\beq
r^{a;b}_{nm}\doteq
\frac{i}{\ww_{nm}}
\left[
  \frac{\vvv^a_{nm}\DDD^b_{nm}+\vvv^b_{nm}\DDD^a_{nm}}{\ww_{nm}}
  -\www^{ab}_{nm}
\right].
\eeq
This expression is symmetric under $a\leftrightarrow b$, and when used
in \eq{I-abc} for $I^{abc}_{mn}$ it renders \eq{sigma-abc} for
$\sigma^{abc}(0;\ww,-\ww)$ totally symmetric, irrespective of crystal
symmetry.  This unphysical behavior is not an artifact of two-band
models, but of the diagonal TBA applied to such models.  The shift
current arising from the photoexcitation of carriers between the two
bands can be calculated exactly, without adding more bands to the
model, by including the additional two-band terms in
\eq{gen-der-wannier} associated with off-diagonal position matrix
elements.  These considerations appear relevant to the ongoing
discussion on the shift-current response of Weyl
semimetals~\cite{osterhoudt-arxiv17,yang-arxiv17,zhang-arxiv18}.

\section{Computational details}
\label{sec:comp-details}

In this section we describe the various steps of the calculations that
we have carried out for two test systems, bulk GaAs and single-layer
GeS. In a first step, we performed density-functional theory
calculations using the {\tt Quantum ESPRESSO} code
package~\cite{gianozzi-jpcm09}. The core-valence interaction was
treated by means of fully-relativistic projector augmented-wave
pseudopotentials (taken from the {\tt Quantum ESPRESSO} website) that
had been generated with the Perdew-Burke-Ernzerhof
exchange-correlation functional~\cite{perdew-prl96}, and the energy
cutoff for the plane-wave basis expansion was set at 60~Ry.
Maximally-localized Wannier functions were then constructed in a
post-processing step, using the {\tt Wannier90} code
package~\cite{wannier90}. Finally, the shift-current spectrum
[\eq{sigma-abc}], the JDOS [\eq{jdos}], and the dielectric function
[\eq{dielectric}] were calculated in the Wannier basis as described in
\sect{wannier}.

\begin{figure}[t]
\includegraphics[width=\columnwidth]{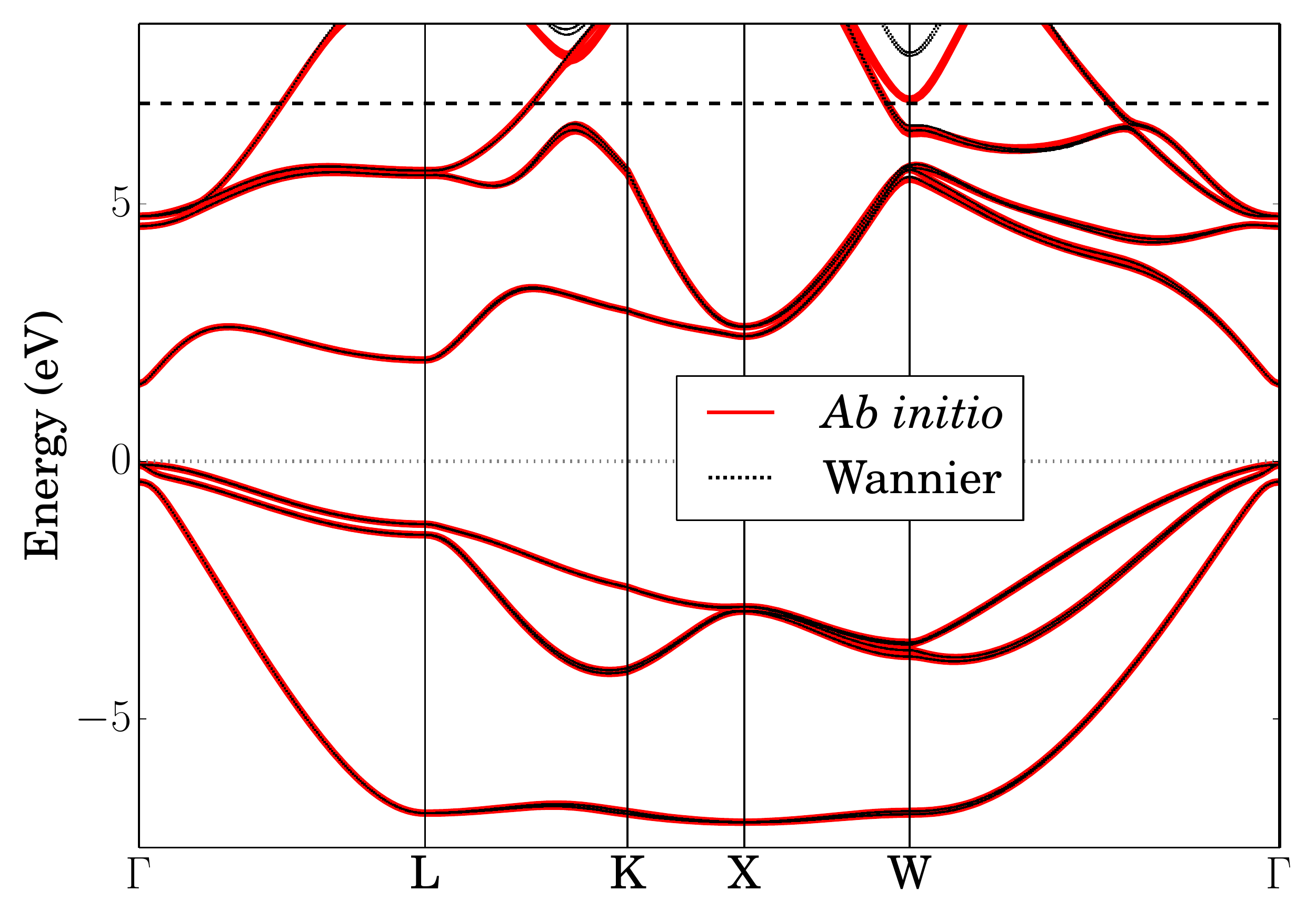}
\caption{The {\it ab initio} and Wannier-interpolated energy bands of
  GaAs, including a scissors correction of 1.15~eV (energies are
  measured from the valence-band maximum). The horizontal dashed line
  at 6.9~eV denotes the upper limit of the inner energy window used in
  the disentanglement step of the Wannier construction procedure.}
\label{fig:1}
\end{figure}

In the case of zincblende GaAs, the self-consistent calculation was
carried out on a $10\times 10\times 10$ \textit{k}-point mesh, using
the experimental lattice constant of $a=10.68\,a_0$.  Starting from
the converged self-consistent Kohn-Sham potential, the 24 lowest bands
and Bloch wavefunctions were then calculated on the same
mesh. Finally, a set of 16 disentangled Wannier functions spanning the
eight valence bands and the eight low-lying conduction bands were
constructed using \textit{s} and \textit{p} atom-centered orbitals as
trial orbitals.  The Wannier-interpolated energy bands are shown in
\fig{1} together with the {\it ab initio} bands (including in both
cases a ``scissors correction'').  The agreement between the two is
excellent inside the inner energy window~\cite{souza-prb01}, which
spans the energy range from the bottom of the figure up to the dashed
horizontal line.

\begin{figure}[t]
  \includegraphics[width=\columnwidth]{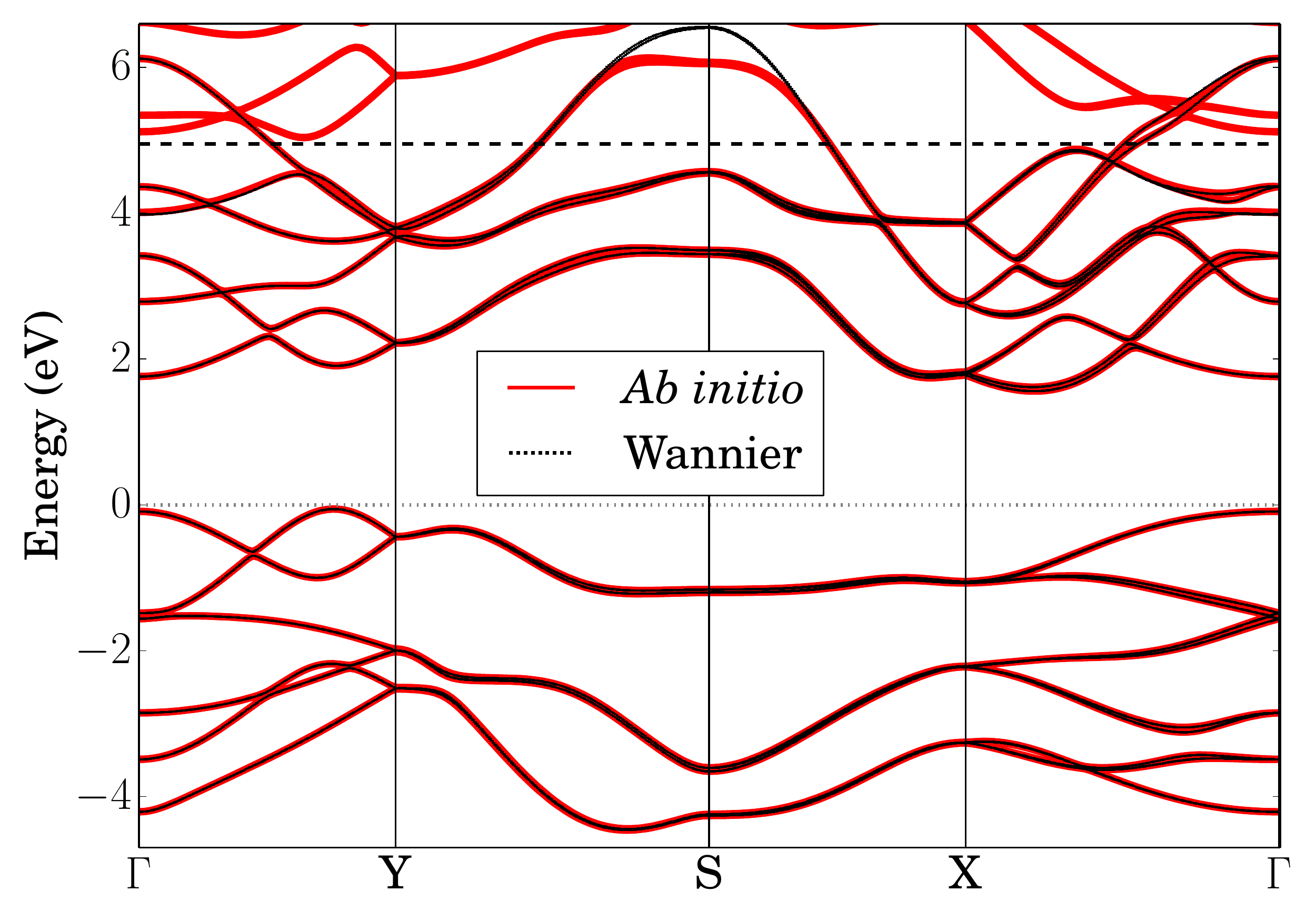}
\caption{The {\it ab initio} and Wannier-interpolated energy bands of
  monolayer GeS. The horizontal dashed line denotes the upper limit of
  the inner energy window.}
\label{fig:2}
\end{figure}

The calculations for monolayer GeS were done in a slab geometry, with
a supercell of length 15~\AA\, along the nonperiodic direction and a
$1\times 12\times 12$ \textit{k}-point mesh for both the
self-consistent and for the band structure calculation. The parameters
for the structure with an in-plane polar distortion were taken from
Table~II in the Supplemental Material of
Ref.~\onlinecite{rangel-prl17}.  Starting from a manifold of 46 bands,
we constructed 32 disentangled Wannier functions spanning the 20
highest valence bands and the 12 lowest conduction bands.  For the
initial projections, we again chose \textit{s} and \textit{p} trial
orbitals centered on each atom.  The {\it ab initio} and
Wannier-interpolated energy bands are shown in \fig{2}.

To obtain well-converged shift-current spectra, we used dense
\textit{k}-point interpolation meshes of $100\times 100\times100$ for
GaAs and $1\times 1000\times 1000$ for GeS.  In the case of GaAs, we
employed an adaptive scheme~\cite{yates-prb07} for choosing the width
of the broadened delta functions in \eq{sigma-abc}.  For GeS we used a
fixed width of 0.02~eV, as it was found to handle better the strong
van-Hove singularities characteristic of two-dimensional (2D) systems.

In the sum-over-states expression for $\sigma^{abc}(0;\ww,-\ww)$, the
energy denominators involving intermediate states should be
interpreted as principal values~\cite{baltz-prb81}.  In our formalism
such denominators appear in
\eqs{gen-der-internal}{gen-der-internal-a}, and in practice we make
the replacement
\beq
\label{eq:eta}
\frac{1}{\ww_{np}}\rightarrow
\frac{\ww_{np}}{\ww_{np}^2+(\eta/\hbar)^2},
\eeq
and similarly for $1/\ww_{pm}$. Such a regularization procedure is
needed to avoid numerical problems caused by near degeneracies.
Following Ref.~\onlinecite{nastos-prb06}, we choose $\eta$ in a range
where the calculated spectrum remains stable.  In the calculations
reported below, we have used $\eta=0.04$~eV for both GaAs and GeS.

As mentioned earlier, a scissors correction was applied to the
calculated band structure of GaAs in \fig{1}, in order to cure the
underestimation of the gap. The conduction bands were rigidly shifted
by 1.15 eV and the spectral quantities plotted in \fig{3} were
modified accordingly as described below, facilitating comparison with
Ref.~\onlinecite{nastos-prb06} where a scissors correction was also
applied.
  
It is clear from \eq{jdos} that the scissors correction leads to a
rigid shift of the JDOS. Although less obvious, the shift-current
spectrum [\eq{sigma-abc}] and the dielectric function
[\eq{dielectric}] also undergo rigid shifts.  The reason is that
\eqs{sigma-abc}{dielectric} do not contain any frequency prefactors,
and the matrix elements therein are intrinsic properties of the Bloch
eigenstates [see \eqs{r}{gen-der}], which are unaffected by the
scissors correction (only the eigenvalues change). The eigenvalues do
appear in \eqs{r-v}{sipe-sum-rule} that are used in practice to
evalute the optical matrix elements, but a careful analysis reveals
that those equations remain invariant under a scissors
correction~\cite{nastos-prb05}.

\section{Results}
\label{sec:results}

\subsection{Bulk GaAs}
\label{sec:gaas}

\begin{figure}[t]
\includegraphics[width=\columnwidth]{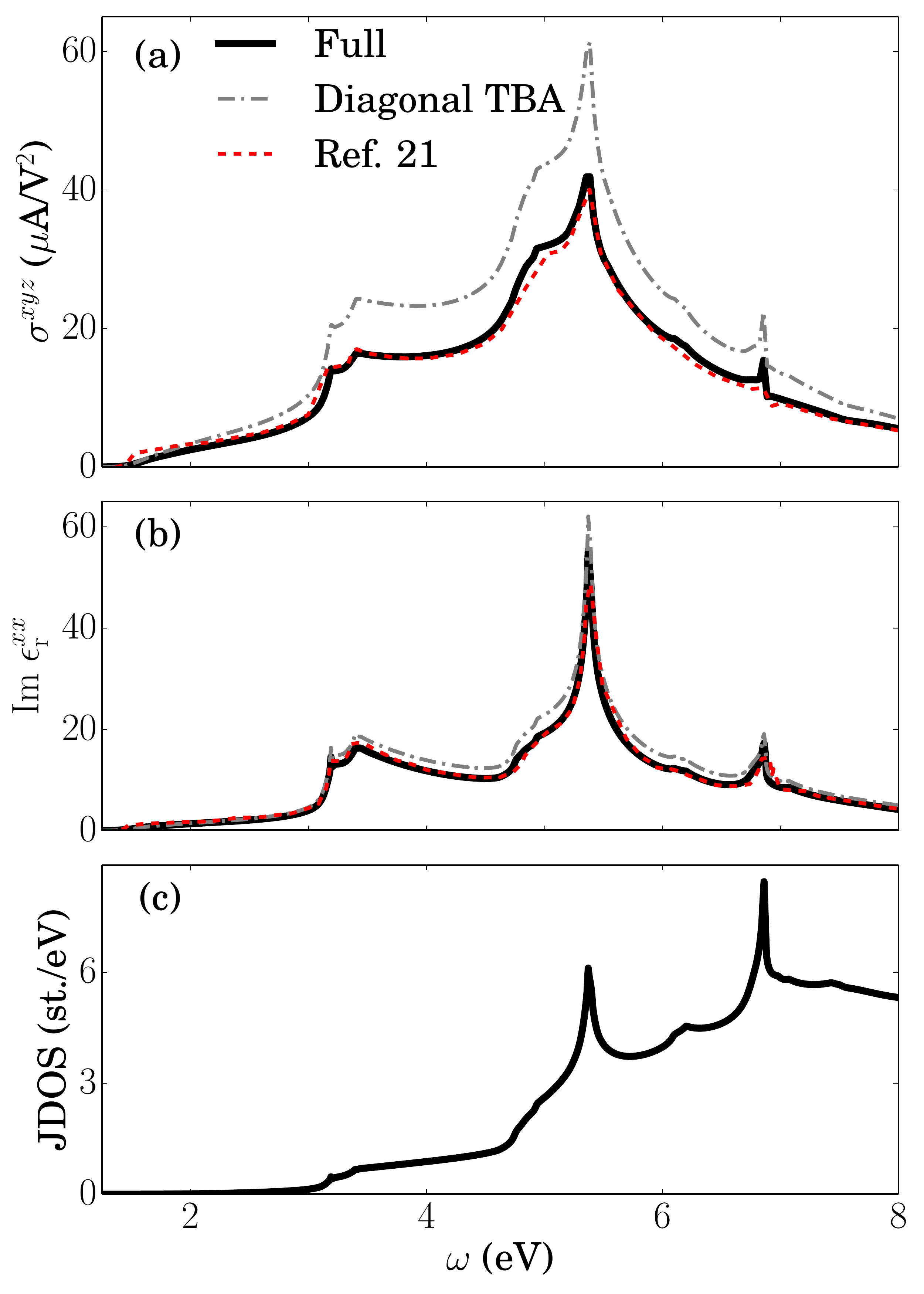}
\caption{(a) Shift-current spectrum, (b) imaginary part of the
  dielectric function, and (c) joint density of states of GaAs,
  calculated by Wannier interpolation including a scissors correction.
  ``Diagonal TBA'' denotes spectra calculated making the diagonal
  tight-binding approximation of \eq{diag} for the optical matrix
  elements. Data adapted from Ref.~\onlinecite{nastos-prb06} is also
  shown.}
\label{fig:3}
\end{figure}

The zincblende semiconductor GaAs was the first piezoelectric crystal
whose shift-current spectrum was evaluated using modern band structure
methods. The original calculation~\cite{sipe-prb00} suffered from a
computational error, and a corrected spectrum was reported
later~\cite{nastos-prb06}. Given the existence of this benchmark
calculation, we have chosen GaAs as the first test case for our
implementation.

Figure~\ref{fig:3}(a) shows the calculated $\sigma^{xyz}(0;\ww,-\ww)$,
which is equal to $\sigma^{abc}(0;\ww,-\ww)$ for any permutation $abc$
of $xyz$, and all other components vanish by
symmetry~\cite{sipe-prb00}.  The imaginary part of the dielectric
function is shown in panel~(b) of the same figure, and the JDOS in
panel~(c).  For comparison, we have included in panels~(a) and~(b) the
spectra calculated in Ref.~\onlinecite{nastos-prb06}.

\begin{figure}[t]
\includegraphics[width=\columnwidth]{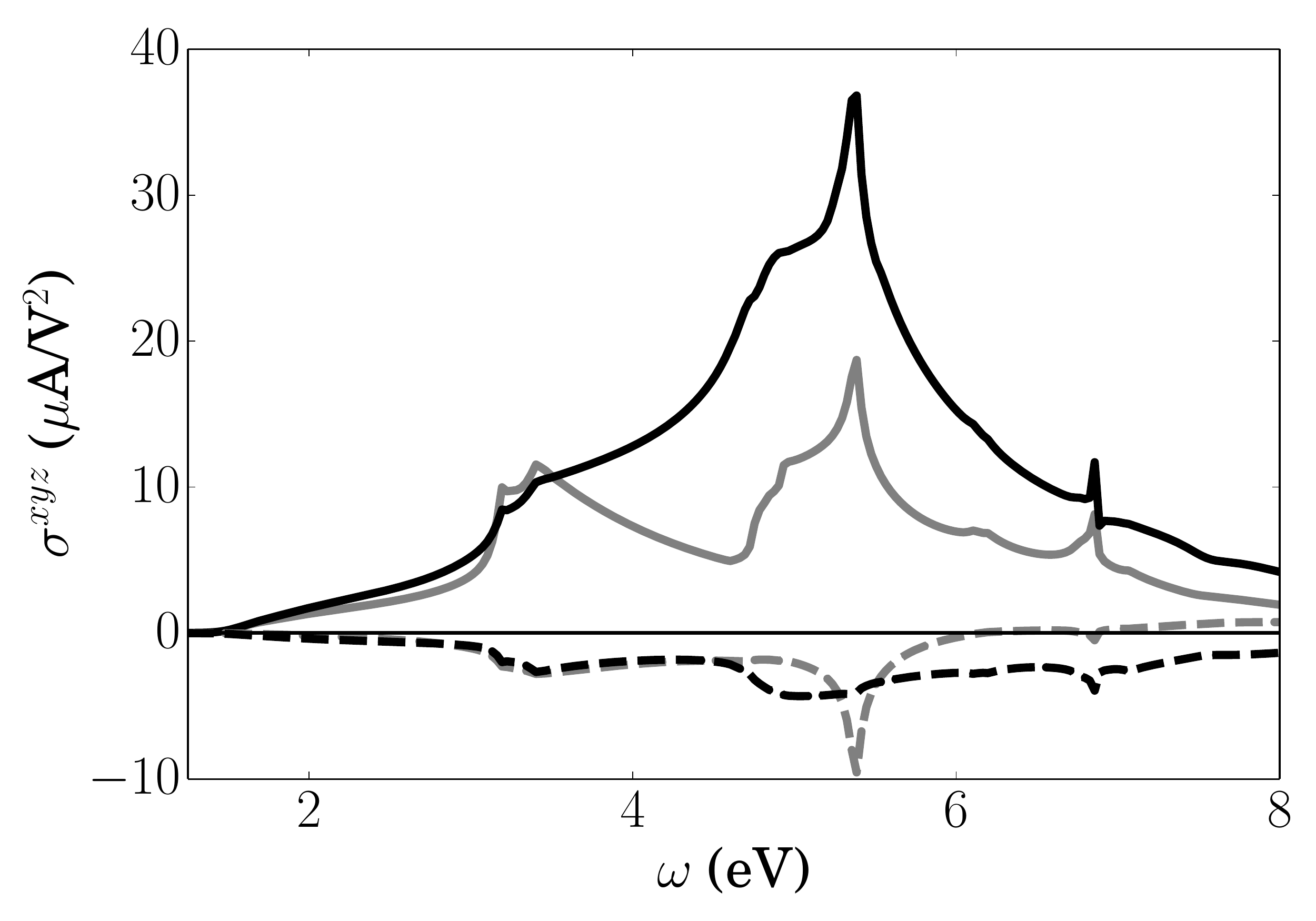}
\caption{ Decomposition of the shift-current spectrum of GaAs shown in
  \fig{3}(a) into ``internal'' (solid lines) and ``external'' (dashed
  lines) terms on one hand, and into ``three-band'' (black lines) and
  ``two-band'' (gray lines) terms on the other.}
\label{fig:4}
\end{figure}

The dielectric function and the shift-current spectrum share similar
peak structures, inherited from the JDOS.  The level of agreement with
Ref.~\onlinecite{nastos-prb06} is excellent for
$\Im\epsilon^{xx}_{\rm r}(\omega)$ and also very good for
$\sigma^{xyz}(0;\ww,-\ww)$, with only minor deviations.  The presence
of small discrepancies is not surprising, given that the shift current
is rather sensitive to the
wavefunctions~\cite{baltz-prb81,belinicher-jetp82,kristoffel-zpb82}
and that the two calculations differ on several technical aspects. For
example, we use pseudopotentials instead of an all-electron method,
and a generalized gradient approximation for the exchange-correlation
potential instead of the local-density approximation.  The BZ
integration methods are also different, and the spin-orbit
contribution to the velocity matrix elements was not included in
Ref.~\onlinecite{nastos-prb06}.

The dash-dotted gray lines in panels~(a) and~(b) of \fig{3} show the
spectra calculated in the diagonal TBA of \eq{diag}.  While in the
case of $\Im\epsilon^{xx}_{\rm r}(\omega)$ the changes are quite
small, they are more significant for $\sigma^{abc}(0;\ww,-\ww)$. This
reflects the strong wave-function dependence of the shift current,
encoded not only in the Wannier centers~\cite{cook-nc17} but also in
the off-diagonal position matrix elements
$\me{\OO n}{\hat\rr}{\RR m}$.  Those matrix elements are usually
discarded in tight-binding calculations, but they should be included
to fully embed the tight-binding model in real space.  The sensitivity
of the shift current to those matrix elements can be understood from
the charge-transfer nature of the photoexcitation process in
piezoelectric
crystals~\cite{baltz-prb81,belinicher-jetp82,nastos-prb06}.

It is instructive to break down the shift-current spectrum calculated
by Wannier interpolation into different types of
contribution. Inserting \eqs{r-wannier}{gen-der-wannier} for
$r^a_{nm}$ and $r^{a;b}_{nm}$ into \eq{I-abc} for $I^{abc}_{nm}$
generates a number of terms. Each can be classified as ``external'' or
``internal'' depending on whether or not it contains off-diagonal
position matrix elements: the term $\rrr^b_{mn}\rrr^{c;a}_{nm}$ is
internal, and all others are external. In addition, we classify each
term as ``two-band'' or ``three-band'' depending on whether it only
involves states $n$ and $m$, or intermediate states $p$ as well. This
gives a total of four types of terms, whose contributions to the shift
current are shown in \fig{4}.

The dominant contribution comes from internal three-band terms, which
by themselves provide a reasonable approximation to the full spectrum
shown in \fig{3}(a).  They are followed by the internal two-band
terms, while the two external terms are somewhat smaller. Over most of
the spectral range, the external terms have the opposite sign compared
to the internal ones. Since the diagonal TBA amounts to discarding the
external terms, that explains why the dash-dotted gray line in
\fig{3}(a) overestimates the magnitude of the full spectrum given by
the solid black line.  We emphasize that the decomposition of the
shift-current spectrum in \fig{4} depends on the choice of Wannier
functions.

\subsection{Monolayer GeS}
\label{sec:ges}

\begin{figure}[t]
  \includegraphics[width=\columnwidth]{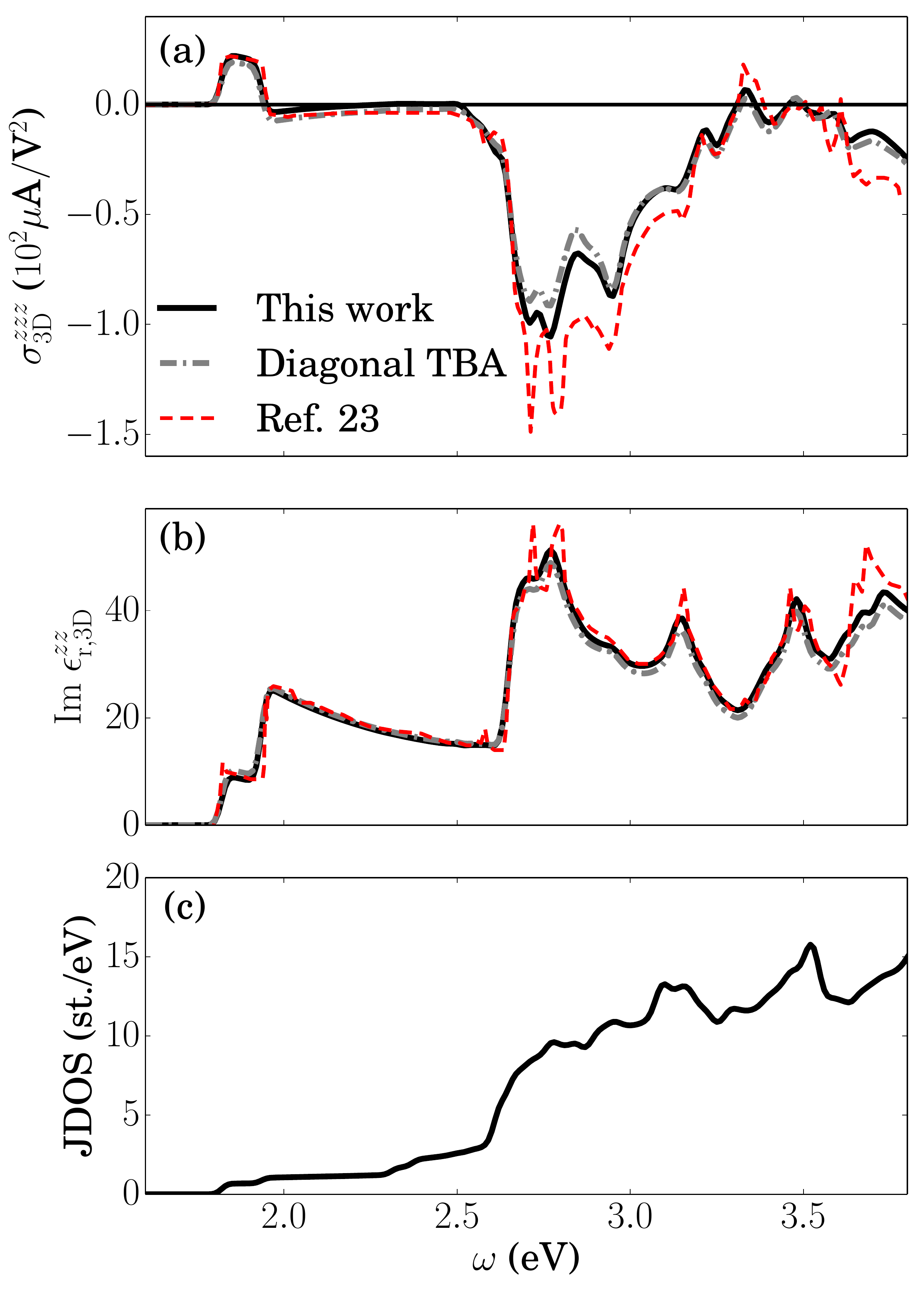}
  \caption{(a), (b) and (c) show the calculated $zzz$ component of the
    shift-current spectrum, the $zz$ component of the dielectric
    function, and the joint density of states of monolayer GeS,
    respectively.  The first two have been rescaled according to
    \eq{3D-slab} to become 3D-like quantities.}
\label{fig:5}
\end{figure}

GeS is a member of the group-IV monochalcogenides, which in bulk form
are centrosymmetric, but become polar --~and hence piezoelectric~--
when synthesized as a single layer. The point group of monolayer GeS
is mm2, which allows for seven tensorial components of
$\sigma^{abc}(0;\ww,-\ww)$ to be nonzero~\cite{rangel-prl17}.  With
the same choice of coordinate axis as in Fig.~1 of
Ref.~\onlinecite{rangel-prl17} (the in-plane directions are
$\hat{\bm y}$ and $\hat{\bm z}$, with the spontaneous polarization
along $\hat{\bm z}$), the nonzero components are $zxx$, $zyy$, $zzz$,
$yyz=yzy$, and $xxz=xzx$.

The $zzz$ component of the shift-current spectrum is displayed in
\fig{5}(a).  Following Ref.~\onlinecite{rangel-prl17}, we report a
3D-like response obtained assuming an active single-layer thickness of
$2.56$\,\AA. This is achieved by rescaling the calculated response of
the slab of thickness $15$\,\AA\, as follows,
\beq
\label{eq:3D-slab}
\sigma^{zzz}_{\rm 3D}=\frac{15}{2.56}\sigma^{zzz}_{\rm slab}.
\eeq
In Figs.~\ref{fig:5}(b,c) we plot the dielectric function [also
rescaled according to \eq{3D-slab}] and the JDOS. As in the case of
GaAs, the main peak structures of the optical spectra in panels~(a)
and~(b) are inherited from the JDOS. The diagonal TBA (dash-dotted
gray lines) changes the calculated spectra only slightly, consistent
with what found in Ref.~\onlinecite{wang-prb17} for monolayer WS$_2$.

Our calculated spectra in \fig{5} are in reasonable agreement with
those reported in Ref.~\onlinecite{rangel-prl17} (dashed red lines),
including on the positions of the main peaks and on the sign change of
the shift current taking place at around 2~eV.  However, the agreement
is not as good as that seen in \fig{3} for GaAs. This may be due in
part to some differences in computational details between the two
calculations, namely the use of different \textit{k}-point meshes and
BZ integration methods: we have sampled the BZ on a uniform mesh of
$10^{6}$ \textit{k} points, while in Ref.~\onlinecite{rangel-prl17} a
more sophisticated tetrahedron method was used for the integration,
but with far fewer \textit{k} points (4900).  There is however another
source of disagreement, which was not present in \fig{3}: the
approximate treatement in Ref.~\onlinecite{rangel-prl17} of the
optical matrix elements within the nonlocal pseudopotential
framework. This source of error is discussed further in
Appendix~\ref{app:nonlocal}.

\subsection{Analysis of computational time}
\label{sec:timing}

Here we compare the computational requirements of our numerical scheme
with a direct calculation of the shift-current spectrum without
Wannier interpolation (e.g., using the method outlined in
Appendix~\ref{app:nonlocal}).  The spectrum is evaluated by
discretizing the BZ integral in \eq{sigma-abc} over a mesh containing
$N$ $k$ points, and we wish to see how the computational times of the
two approaches scale with $N$.

For that purpose, let us define the following time scales per $k$
point: $t_{\rm w}$ and $t_{\rm d}$ are the times to evaluate the
integrand in \eq{sigma-abc} by Wannier interpolation and using the
direct method, respectively, and $t_{\rm nscf}$ is the time to carry
out a non-self-consistent calculation to obtain the \textit{ab initio}
Bloch eigenfunctions and energy eigenvalues.  Further, we define
$T_{\rm scf}$ as the total time needed to carry out the
self-consistent ground-state calculation, and $T_{\rm wf}$ as the
total time needed to construct the Wannier functions on a grid of $M$
$k$ points.  The total time of a Wannier-based calculation of the
shift current is then
\beq
\label{eq:T-wannier}
T_{\rm scf}+Mt_{\rm nscf}+T_{\rm wf}+Nt_{\rm w},
\eeq
while the total time of a direct calculation is
\beq
\label{eq:T-direct}
T_{\rm scf}+N\left(t_{\rm nscf}+t_{\rm d}\right)\approx
T_{\rm scf}+Nt_{\rm nscf},
\eeq 
where we used $t_{\rm d}\ll t_{\rm nscf}$.

\begin{figure}[t]
  \includegraphics[width=\columnwidth]{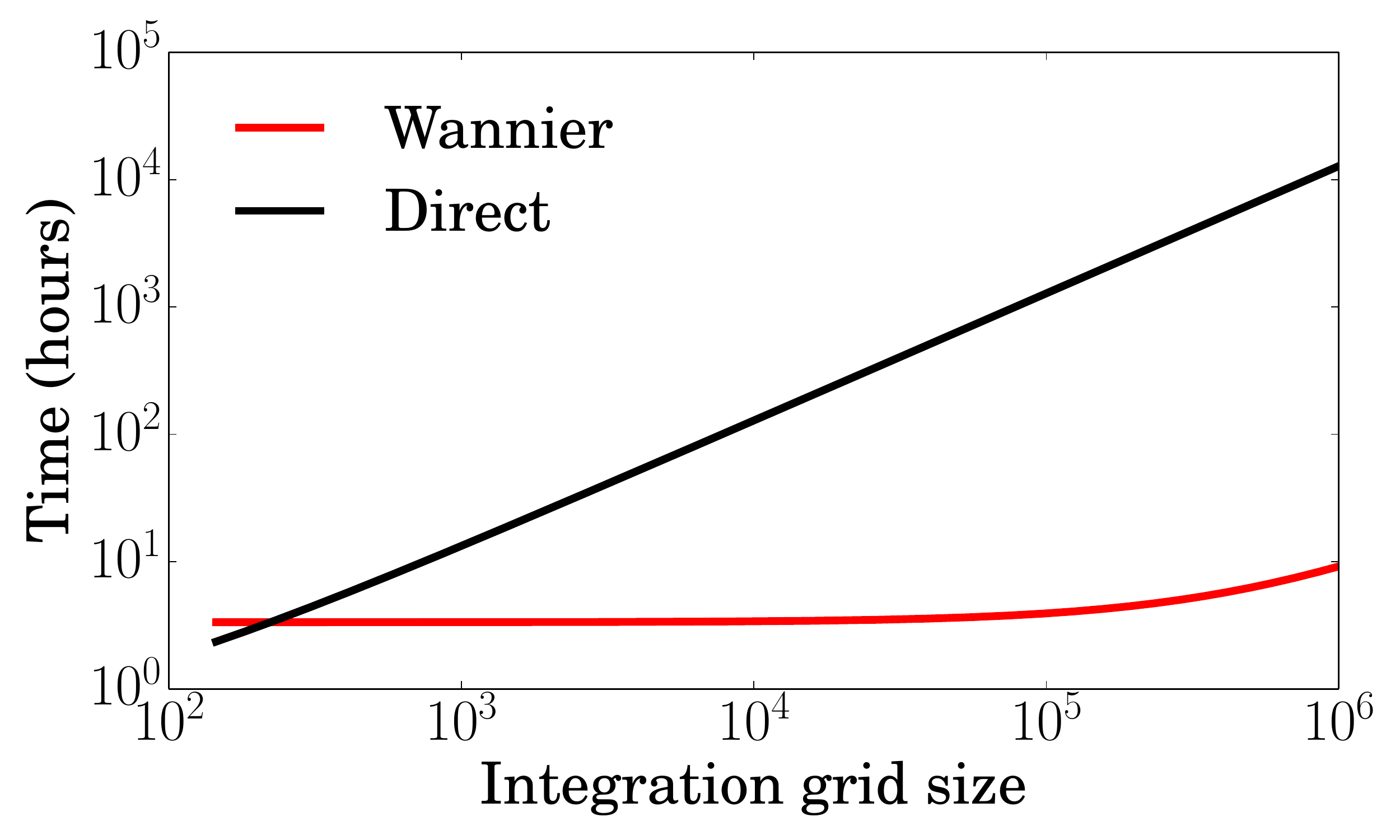}
  \caption{Time estimates for calculating the shift-current spectrum
    of monolayer GeS on a single processor with and without Wannier
    interpolation [\eqs{T-wannier}{T-direct}, respectively], as a
    function of the size $N$ of the BZ integration grid.}
\label{fig:6}
\end{figure}

Let us take as a concrete example a calculation for monolayer GeS done
on a single Intel Xeon E5-2680 processor with 24 cores running at
2.5~GHz. For the choice of parameters indicated in \sect{comp-details}
we find $t_{\rm w}\simeq 21$~ms, $t_{\rm nscf}\simeq 46$~s,
$T_{\rm scf}\simeq 0.5$~hours, and $T_{\rm wf}\simeq 1$~hour.  In
\fig{6} we plot as a function of $N$ the total times obtained from
\eqs{T-wannier}{T-direct}, for $M=12^2$. The use of Wannier
interpolation is already quite advantageous for $N\sim 500$, and the
speedup increases very rapidly with $N$. If a dense $k$-point sampling
with $N\sim 10^6$ is required, the speedup reaches three orders of
magnitude. (The absolute times reported in \fig{6} can be reduced by
parallelizing the loop over the $N$ $k$~points, which is trivial to do
both with and without Wannier interpolation.)

\section{Summary}
\label{sec:summary}

In summary, we have described and validated a Wannier-interpolation
scheme for calculating the shift-current spectrum of piezoelectric
crystals, starting from the output of a conventional
electronic-structure calculation. The method is both accurate and
efficient; this is achieved by using a truncated Wannier-function
basis, but without incurring in truncation errors when evaluating the
optical matrix elements.  The same approach can be applied to other
nonlinear optical responses, such as second-harmonic generation, that
involve the same matrix elements~\cite{sipe-prb00,wang-prb17}.

Our work was motivated in part by the growing interest in the
calculation of nonlinear optical properties of novel materials such as
Weyl semimetals and 2D materials. We hope that the proposed
methodology, and its implementation in the {\tt Wannier90} code
package, will help turn such calculations into a fairly routine task.

When describing the formalism, we tried to emphasize the notion that
Wannier functions provide an essentially exact (in some chosen energy
range) tight-binding parametrization of the {\it ab initio} electronic
structure. Thus, we chose our notation and conventions so as to
facilitate comparison with the expressions for nonlinear optical
responses found in the tight-binding literature.  Our numerical
results suggest that it should be possible to systematically improve
the tight-binding description of such responses by including
off-diagonal position matrix elements as additional model parameters.
In Ref.~\onlinecite{bennetto-prb96}, an attempt was made along those
lines to improve the tight-binding parametrization of semiconductors
for the calculation of Born effective charges, but with limited
success.  Clearly more work is needed in this direction, and the shift
current, with its strong sensitivity to the wavefunctions, is
particularly well-suited for such investigations.

\begin{acknowledgments}

  The authors gratefully acknowledge stimulating discussions with
  Fernando de Juan, Jianpeng Liu, Cheol-Hwan Park, and David
  Vanderbilt. They also thank Cheol-Hwan Park for bringing
  Ref.~\onlinecite{wang-prb17} to their attention when the work was
  close to completion, and Chong Wang for elucidating the relation
  between the formalism of Ref.~\onlinecite{wang-prb17} and the one
  presented here. The work was supported by Grant No.~FIS2016-77188-P
  from the Spanish Ministerio de Econom\'ia y Competitividad, and by
  Elkartek Grant No. KK-2016/00025.  Computing time was granted by the
  JARA-HPC Vergabegremium and provided on the JARA-HPC Partition part
  of the supercomputer JURECA at Forschungszentrum J\"ulich.

\end{acknowledgments}

\appendix

\section{Symmetry considerations}
\label{app:symm}

As mentioned in the Introduction, the shift current vanishes in
centrosymmetric crystals. To verify that \eq{sigma-abc} behaves
correctly in that limit, note that the presence of inversion symmetry
implies the relations
\begin{subequations}
\begin{align}
E_n(-\kk)&=E_n(\kk),\\
I^{abc}_{mn}(-\kk)&=-I^{abc}_{mn}(\kk).
\end{align}
\end{subequations}
Hence $\kk$ and $-\kk$ give equal and opposite contributions to the BZ
integral in \eq{sigma-abc}, leading to $\sigma^{abc}(0;\ww,-\ww)=0$.

The shift current has been mostly studied in acentric crystals without
magnetic order. The presence of time-reversal symmetry in such systems
implies
\begin{subequations}
\begin{align}
E_n(-\kk)&=E_n(\kk),\\
I^{abc}_{mn}(-\kk)&=-\left[I^{abc}_{mn}(\kk)\right]^*.
\end{align}
\end{subequations}
The points $\kk$ and $-\kk$ now give equal contributions to the BZ
integral, and \eq{sigma-abc} reduces to
\bea
\label{eq:sigma-abc-TR}
\sigma^{abc}(0;\ww,-\ww)&=&-\frac{i\pi e^3}{2\hbar^2}
\int\dk\sum_{n,m}f_{nm}
\left(I^{abc}_{mn}+I^{acb}_{mn}\right)\nn
&\times&\delta(\omega_{mn}-\omega),
\eea
which is Eq.~(57) in Ref.~\onlinecite{sipe-prb00}.  For $b=c$, this
form remains equivalent to \eq{sigma-abc} even without time-reversal
symmetry.

\section{Comparison with Ref.~\onlinecite{wang-prb17}}
\label{app:comparison}

In Ref.~\onlinecite{wang-prb17}, a similar Wannier-interpolation
scheme for calculating the shift current was proposed independently.
The expression given in that work for the generalized derivative in
the Wannier basis is however different from \eq{gen-der-wannier}.  In
this Appendix, we show that the two formulations are in fact
consistent with one another.

Below their Eq.~(7), the authors of Ref.~\onlinecite{wang-prb17} write
\bea
\label{eq:dbAa}
\partial_bA^a&=&\left(\partial_b U^\dagger\right)A_a\W U
+U^\dagger\left(\partial_bA_a\W\right)U\nn
&+&U^\dagger A_a\W\partial_bU
+i\left(\partial_b U^\dagger\right)\partial_a U
+iU^\dagger\partial^2_{ab}U,
\eea
which follows from differentiating \eq{A-wannier}.  The last term can
be expressed in terms of $D^a=U^\dagger\partial_a U=-i\AAA^a$ as
\beq
\label{eq:d2abD}
iU^\dagger\partial^2_{ab}U=i\partial_b D^a+iD^bD^a.
\eeq
The non-Hermitean term $iD^bD^a$ cancels the fourth term in \eq{dbAa},
leaving an expression for $\partial_bA^a$ that is correctly Hermitean,
term by term. Let us now evaluate the term $\partial_b D^a$ assuming
$D^a_{nn}=0$ (parallel-transport)~\cite{wang-prb17}. The off-diagonal
matrix elements of the matrix $D^a$ read
\beq
D^a_{nm}=-\frac{\vvv^a_{nm}}{\ww_{nm}}\quad(m\not=n),
\eeq
where 
$\vvv^a_{mm}$ was defined in \eq{vvv}.
Invoking \eq{dbObar} we find
\bea
\label{eq:comparison}
\partial_bD^a_{nm}&=&-\frac{1}{\ww_{nm}}
\Big(
  \www^{ab}_{nm}
  -\sum_{l\not= m}\,\frac{\vvv^a_{nl}\vvv^b_{lm}}{\ww_{lm}}
  -\sum_{l\not= n}\,\frac{\vvv^b_{nl}\vvv^a_{lm}}{\ww_{ln}}\nn
  &+&\frac{\vvv^b_{mm}\vvv^a_{nm}}{\ww_{nm}}
  -\frac{\vvv^b_{nn}\vvv^a_{nm}}{\ww_{nm}}
\Big)\quad(m\not=n),
\eea
with $\www^{ab}_{nm}$ given by \eq{www}.  Substituting the term
$\partial_br^a_{nm}$ in \eq{gen-der} by \eq{dbAa} combined with
\eqs{d2abD}{comparison}, \eq{gen-der-wannier} for $r^{a;b}_{nm}$ is
eventually recovered (after using $\AAA^b=\rrr^b$, which holds in a
parallel-transport gauge).

We can now proceed to compare with
Ref.~\onlinecite{wang-prb17}. Combining
Eqs.~(\ref{eq:d2abD})--(\ref{eq:comparison}) we obtain
\begin{widetext}
\bea
\label{eq:d2abU}
\left(U^\dagger\partial^2_{ab}U\right)_{nm}
&=&-\frac{1}{\ww_{nm}}
\left(
  \www^{ab}_{nm}
  -\sum_{l\not= m}\,\frac{\vvv^a_{nl}\vvv^b_{lm}}{\ww_{lm}}
  -\sum_{l\not= n}\,\frac{\vvv^b_{nl}\vvv^a_{lm}}{\ww_{ln}}
  +\frac{\vvv^b_{mm}\vvv^a_{nm}}{\ww_{nm}}
-\frac{\vvv^b_{nn}\vvv^a_{nm}}{\ww_{nm}}\right)
+\sum_{l\not=n,m}\,\frac{\vvv^b_{nl}}{\ww_{nl}}
\frac{\vvv^a_{lm}}{\ww_{lm}}.
\eea
\end{widetext}
The first two terms in this equation agree with those in Eq.~(8) of
Ref.~\onlinecite{wang-prb17}, and in the following we show that the
remaining terms in both equations can also be brought into agreement.
Dropping the first two terms of \eq{d2abU} and using
$\ww_{nm}/(\ww_{nl}\ww_{lm})=1/\ww_{nl}-1/\ww_{lm}$ in the last term,
we find\footnote{\equ{idents-wang} was obtained by Chong Wang,
  commenting on an earlier version of the manuscript (private
  communication).}
\begin{widetext}
\bea
\label{eq:idents-wang}
&&-\frac{1}{\ww_{nm}}
\left(
  -\sum_{l\not=n}\,\frac{\vvv^b_{nl}\vvv^a_{lm}}{\ww_{ln}}
  +\frac{\vvv^b_{mm}\vvv^a_{nm}}{\ww_{nm}}
  -\frac{\vvv^b_{nn}\vvv^a_{nm}}{\ww_{nm}}
  -\sum_{l\not=n,m}\,\frac{\vvv^b_{nl}\vvv^a_{lm}}{\ww_{nl}}
  -\sum_{l\not=n,m}\,\frac{\vvv^b_{nl}\vvv^a_{lm}}{\ww_{lm}}
\right)\nn
&=&-\frac{1}{\ww_{nm}}
\left(
  -\sum_{l\not=n,m}\,\frac{\vvv^b_{nl}\vvv^a_{lm}}{\ww_{lm}}
  +\frac{\vvv^b_{nm}\vvv^a_{mm}}{\ww_{nm}}
  +\frac{\vvv^b_{mm}\vvv^a_{nm}}{\ww_{nm}}
  -\frac{\vvv^b_{nn}\vvv^a_{nm}}{\ww_{nm}}
\right)\nn
&=&-\frac{1}{\ww_{nm}}
\left(
  -\sum_{l\not=m}\,\frac{\vvv^b_{nl}\vvv^a_{lm}}{\ww_{lm}}
  +\frac{\vvv^b_{nm}\vvv^a_{mm}}{\ww_{nm}}
  +\frac{\vvv^b_{mm}\vvv^a_{nm}}{\ww_{nm}}
\right),
\eea
\end{widetext}
which is indeed identical to the last three terms in Eq.~(8) of
Ref.~\onlinecite{wang-prb17}. It is worth mentioning that in this
formulation the Hermiticity of $r^{a;b}_{nm}$ is only satisfied
globally, not term by term as in the case of \eq{gen-der-wannier}.

\section{Berry curvature in the Wannier basis:
Removal of the parallel-transport assumption}
\label{app:curv}

In Ref.~\onlinecite{wang-prb06}, around Eqs.~(23)--(24), a
parallel-transport gauge was imposed on the $U$ matrices while
evaluating the Berry curvature in a Wannier basis.  Should one then
enforce the parallel-transport condition when choosing those matrices
at neighboring $k$~points?  This is in fact not necessary, as we now
show.

The Berry curvature of band~$n$ is given by the $m=n$ element of the
matrix
\beq
\label{eq:curv}
\Omega^{ab}_{\kk nm}=
i\ip{\partial_a u_{\kk n}}{\partial_b u_{\kk m}}-
i\ip{\partial_b u_{\kk n}}{\partial_a u_{\kk m}}.
\eeq
Using
\beq
\ket{\partial_a u_n}=\sum_j\,\ket{\partial_a u\W_j}U_{jn}-i\sum_m\,
\ket{u_m}\AAA^a_{mn},
\eeq
which follows from \eqs{u-H}{delU}, we find
\beq
\label{eq:Omega}
\Omega^{ab}=\overline\Omega_{ab}+i\left[\AAA^a,\overline A_b\right]
-i\left[\AAA^b,\overline A_a\right]+i\left[\AAA^a,\AAA^b\right].
\eeq
This is Eq.~(27) of Ref.~\onlinecite{wang-prb06}, in a slightly
different notation.  Recall from \eq{AAA} that $\AAA^a$ is the Berry
connection for the $U$ matrices; instead of imposing the
parallel-transport condition $\AAA^a_{nn}=0$ as done in
Ref.~\onlinecite{wang-prb06}, we let $\AAA^a_{nn}$ be nonzero and
write $\AAA^a_{nm}=\delta_{nm}\AAA^a_{nn}+\rrr^a_{nm}$, in accordance
with \eq{rrr}.  The first commutator in \eq{Omega}, for example,
becomes
\beq
i\left[\rrr^a,\overline A_b\right]_{nm}-
i\overline A_{b,nm}\left(\AAA^a_{mm}-\AAA^a_{nn}\right).
\eeq
Since the second term vanishes for $m=n$, we conclude that the Berry
curvature, given by the band-diagonal entries in \eq{Omega}, is
insensitive to the value of the gauge-dependent quantity
$\AAA^a_{nn}$. This is consistent with the fact that the Berry
curvature is gauge invariant.

\section{Approximate treatment of the optical matrix elements with
  nonlocal pseudopotentials}
\label{app:nonlocal}

In some previous {\it ab initio} calculations of the shift
current~\cite{nastos-prb06,rangel-prl17}, the velocity operator was
approximated as
\beq
\label{eq:v-p}
 \hat{\bm v}=\frac{\hat{\bm p}}{m_e}
=-\frac{i\hbar}{m_e}{\bm\nabla}_\rr.
\eeq
The interband velocity matrix elements $v_{nm}$ in the Bloch basis
were then inserted into \eqs{r-v}{sipe-sum-rule} (dropping the term
$w^{ab}_{nm}$ in the latter) to obtain the interband dipole matrix
$r^a_{nm}$ and its generalized derivative $r^{a;b}_{nm}$.

When using either an all-electron method (as in the GaAs calculation
of Ref.~\onlinecite{nastos-prb06}) or local pseudopotentials, the
above procedure is exact, at least when spin-orbit coupling is
neglected.\footnote{The spin-orbit-interaction gives an additional
  contribution to the velocity operator~\cite{blount}.  That
  contribution is typically small and can be safely neglected, as done
  in Ref.~\onlinecite{nastos-prb06}. In our formulation, that
  contribution is automatically included.} However, modern
pseudopotential calculations employ nonlocal pseudopotentials, for
which that procedure introduces some errors: the velocity operator is
not simply given by \eq{v-p}~\cite{baroni-prb86,hybertsen-prb87}, and
as a result the term $w^{ab}_{nm}$ in \eq{sipe-sum-rule} for
$r^{a;b}_{nm}$ becomes nonzero (see Appendix~B in
Ref.~\onlinecite{wang-prb17}).

In this Appendix we perform additional calculations for single-layer
GeS employing the same computational setup as used in
Ref.~\onlinecite{rangel-prl17} ({\tt ABINIT} code~\cite{gonze-cpc09}
with Hartwigsen-Goedecker-Hutter
pseudopotentials~\cite{hartwigsen-prb98}), in order to estimate the
errors arising from the use of the approximate procedure outlined
above.

\begin{figure}
\includegraphics[width=\columnwidth]{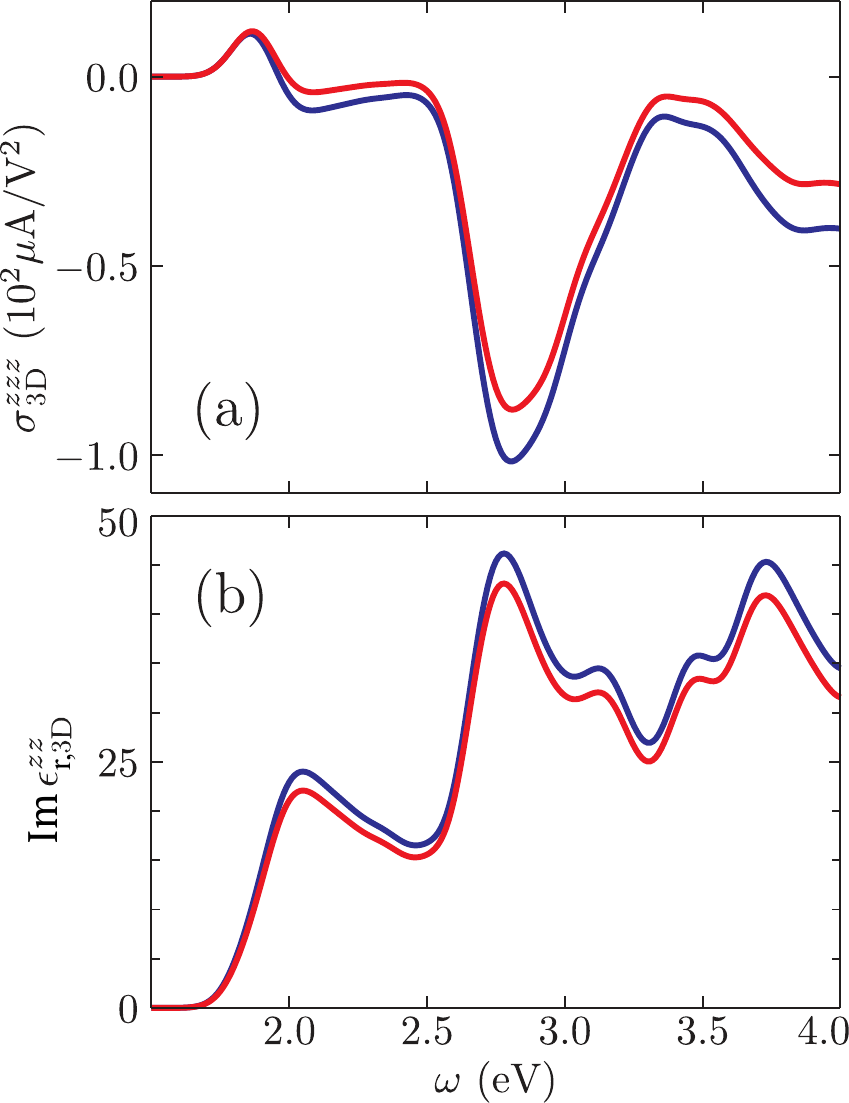}
\caption{\label{fig:wanVSabinit} (a) Shift-current spectrum, and (b)
  dielectric function of single-layer GeS calculated using an exact
  (red) and an approximate (blue) treatment of the optical matrix
  elements within the nonlocal-pseudopotential approach.  The red
  curve was obtained with Wannier interpolation, while for the blue
  curve the optical matrix elements were calculated directly in the
  plane-wave basis using \eq{v-p}.}
\label{fig:7}
\end{figure}

As a first step, we switched off by hand the nonlocal terms in the
pseudopotentials. For a given $k$-point sampling and delta-function
smearing, the resulting spectra $\Im\epsilon^{zz}_{\rm r}(\ww)$ and
$\sigma^{zzz}(0;\ww,-\ww)$ (not shown) were found to be in perfect
agreement with those calculated by Wannier interpolated using the same
local pseudopotentials.  This provided a strong numerical check of our
Wannier interpolation scheme, which does not depend on whether an
all-electron or a pseudopotential method has been used, or on whether
the pseudopotentials are local or nonlocal.

We then redid both calculations using the full nonlocal
pseudopotentials.  The results obtained by sampling the 2D BZ on a
relatively coarse $70\times 70$ grid with a fairly large
delta-function broadening of 0.1~eV are shown in \fig{7} (as a result
of the coarse $k$-point sampling and of the large broadening, the
spectral features are broadened compared to \fig{5}).  There are clear
differences between the spectra calculated in the manner of
Ref.~\onlinecite{rangel-prl17}, and those obtained using the Wannier
interpolation scheme: the positions of the peaks are the same, but
their heights are somewhat different, as expected from a small change
in the matrix elements.  Given the perfect agreement that had been
found with local pseudopotentials, these differences must arise
exclusively from the approximate treatment of the optical matrix
elements in the approach of Ref.~\onlinecite{rangel-prl17} combined
with nonlocal pseudopotentials. Since the level of disagreement seen
in \fig{7} is comparable to that seen in Figs.~\ref{fig:5}(a,b), it
seems plausible that there the discrepancies may also arise in part
from these small errors in the matrix elements.

\bibliography{biblio}

\end{document}